# TECHNICAL SOLUTIONS TO RESOURCES ALLOCATION FOR DISTRIBUTED VIRTUAL MACHINE SYSTEMS

Ha Huy Cuong Nguyen
Departmentof
InformationTechnology,
Quangnam University
Quang Nam, Viet Nam
nguyenhahuycuong@gmail.com

Van Thuan Dang
Departmentof
InformationTechnology,
Industrial University of HCM City
Quang Ngai, Viet Nam
dangvanthuan26@gmail.com

Van Son Le
Departmentof
InformationTechnology,
Danang University of Education
The university of Danang
Da Nang, Viet Nam
levansupham2004@yahoo

*Abstract*— Virtual machine is built on group of real servers which are scattered globally and connect together through the telecommunications systems, it has an increasingly important role in the operation, providing the ability to exploit virtual resources. The latest technique helps to use computing resources more effectively and has many benefits, such as cost reduction of power, cooling and, hence, contributes to the Green Computing. To ensure the supply of these resources to demand processes correctly and promptly, avoiding any duplication or conflict, especially remote resources, it is necessary to study and propose a reliable solution appropricate to be the foundation for internal control systems in the cloud. In the scope of this paper, we find a way to produce efficient distributed resources which emphasizes solutions preventing deadlock and proposing methods to avoid resource shortage issue. With this approach, the outcome result is the checklist of resources state which has the possibility of deadlock and lack of resources, by sending messages to the servers, the server would know the situation and have corresponding reaction.

I. INTRODUCTION

In the early 1980s, Cloud Computing (Clouds) has changed from large computer models to client - server model. Details infrastructure is abstracted from the users, they do not need to know about IT infrastructure and resources are easily accessible in the cloud. Client use of cloud computing applications while computing resources or data placed in the cloud environment. Most cloud computing infrastructure consists of services delivered through data centers and built on the virtual machine. Cloud computing resources are often a single points of access to all cloud computing servers. At the moment, the Internet retains its traditional role as a means of communication and at the same time, it is also a means to share resources. The current trend shows the need to build more flexibility infrastructure in scalability, resilience of security and network congestion. Virtualization technology provides the abstract and isolates the lower level functions, allowing greater mobility and gathering physical resources [2].

These problems have prompted researchers, expertist in the field of computer science looking for better solutions to meet capacity requirements of information technology service from users. In this article, we present solutions to virtual machine model which needs to provide information resources, preventing deadlock in resources supply. Deadlock problems in resources supply on distributed platforms has always been an interest of advanced researchers. However, there are still many things to do with the challenge of future trends.

In the past, grid computing and batch scheduling have both been commonly used for large scale computation. Cloud computing presents a different resource allocation paradigm than either grids or batch schedulers [4,5]. In particular, Amazon C2 [10], is equipped to, handle may smaller computer resource allocations, rather than a few, large request as is normally the case with grid computing. The introduction of heterogeneity allows clouds to be competitive with traditional distributed computing systems, which often consist of various types of architecture as well. In a heterogeneous cloud environment. Recently, reports have appeared many of the studies provide cloud computing resources, the majority of this research to deal with variability in resource capacity for infrastructure and application performance in the cloud. In this paper, we develop a method to predict the lease completion time distribution that is applicable to making a sophisticated trade off decisions in resource allocation and scheduling. Our evaluation shows that these methods deadlock detection using algorithm two ways search can improve efficiency and effectiveness of the cloud computing allocation resource heterogeneous systems.

The work is organized in the following way: in section 2, we introduce the related works; in section 3, we introduce existing models; in section 4, we present solutions in resource allocation heterogeneous distributed vitual machine, in section 5, we present the results from our assessment in section 6, we present our conclusions and suggestions for future work.

II. RELATED WORKS

Large distributed system [6.10] using the virtualization technology to enable the creation of dynamic range of virtual resources which can meet the computing needs of users with specific applications, grid computing.

Resource allocation in cloud computing has attracted the attention of the research community over last few years. Srikantaiah et al. [8] studied the problem of request scheduling for multi-tiered web applications in virtualized heterogeneous systems in order to minimize energy consumption while meeting performance requirements. They proposed a heuristic for a multidimensional packing problem as an algorithm for workload consolidation. Garg et al. [10] proposed near optimal scheduling policies that consider a number of energy efficiency factors, which changes across different data centers depending on their location, architectural design, and management system. Warneke et al. [11] discussed the challenges and opportunities for efficient parallel data processing in cloud environment and presented a data processing framework to exploit the dynamic resource provisioning offered by IaaS clouds. Wu et al. [12] proposed a resource allocation for SaaS providers who want to minimize infrastructure cost and SLA violations. Addis et al. [13] proposed resource allocation policies for the management of multi-tier virtualized cloud systems with the aim to maximize the profits associated with multiple class SLAs. A heuristic solution based on a local search that also provides availability, guarantees that running applications has been developed.

Distributed intelligent model has been proposed to support for large complex distributed systems with smart algorithms. The concept of distributed intelligence model aims to provide information resources based on middleware components which can meet the growing and challenging request from customer that they do not necessarily have to change the system.

The trend in building virtual machine network model in order to manage resources effectively include the following useful purposes:
- The hardware resources in distributed system consists of separate compute nodes connected together via communications networks. At each node, resources include CPU, memory, disk, network, computers, clusters, grids. The special thing is, it cannot communicate directly to the resources of other nodes at this node. The physical architecture components can be the same or different ... These buttons can be distributed on any geographical surface and in separate governance areas, management by the resource management system.
- For resources information including system of programs and data, a vital key requirement of the system is to ensure the coherence of data in multiple host systems.

The process of resource providing under virtualization mechanism is illustrated in Figure 1. Grouping the cloud computing service providers activate based on the need for additional resources and the need for collaboration which has been explained in the basic functions of the cloud computing architecture, and in the structure of the cloud.

Multi-agent system in resources providing based on virtualization mechanism

### III. SYSTEM MODEL RESOURCE ALLOCATION IN HETEROGENEOUS DISTRIBUTED PLATFORMS

Resource allocation in cloud computing has attracted the attention of the research community in the last few years. Cloud computing presents a different resource allocation paradigm than either grids or batch schedulers[2]. In particular, Amazon C2 [10], is equipped to, handle may smaller computer resource allocations, rather than a few, large request as is normally the case with grid computing. The introduction of heterogeneity allows clouds to be competitive with traditional distributed computing systems, which often consist of various types of architecture as well.

Like traditional distributed system before we can see a heterogeneous distributed system consists of a set of processes that are connected by a communication network. The communication delay is finite but unpredictable [21,22].

#### A. The application

A heterogeneous distributed program is composed of a set of n asynchronous processes $p_1, p_2, \ldots, p_n$ that communicates by message passing over the communication network. We assume that each process is running on a different processor. The processor does not share a common global memory and communicate solely by passing messages over the communication network. There is no physical global clock in the system to which processes have instaneous access. The communication medium may deliver messages out of order, messages may be lost garble or duplicated due to timeout and retransmission, processors may fail and communication links may go down. The system can be modeled as a directed graph in which vertices represent the processes and edge represent unidirectional communication channels.

**Example 1** Resource allocation on heterogeneous distributed platforms

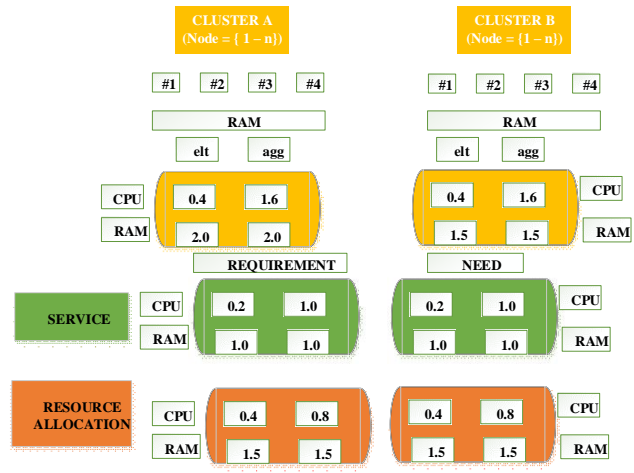

Figure 1. Example problem instance with two nodes and one service, showing possible resource allocations.

Figure 1 illustrates an example with two nodes and one service. Node A, B are comprised of 4 cores and a large memory. Its resource capacity vectors show that each core has elementary capacity 0.8 for an aggregate capacity of 3.2. Its memory has a capacity of 1.0, with no difference between elementary and aggregate values because the memory, unlike cores, can be partitioned arbitrarily. No single virtual CPU can run at the 0.9 CPU capacity on this node. The figure shows two resource allocations one on each node. On both nodes, the service can be allocated for memory it requires.

Informally speaking, a deadlock is a system state where requests are waiting for resources held by other requesters which, in turn, are also waiting for some resources held by the previous requests. In this paper, we only consider the case where requests are processors on virtual machine resource allocation on heterogeneous distributed platforms. A deadlock situation results in permanently blocking a set of processors from doing any useful work.

There are four necessary conditions which allow a system to deadlock[3]: (a) Non – Preemptive: resources can only be released by the holding processor; (b) Mutual Exclusion: resources can only be accessed by one processor at a time; (c) Blocked Waiting: a processor is blocked until the resource becomes available; and (d) Hold – and – Wait: a processor is using resources and making new requests for other resources that the same time, without releasing held resources until some time after the new requests are granted.

**Example 2** A example simple platform

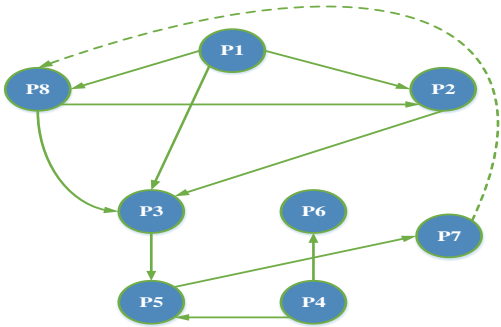

Figure 2.  A example simple platform

We use the platform graph, for the grid platform. We model a collection of heterogeneous resources and the communication links between them as the nodes and edges of an undirected graph. Seea example in Figure 2 with 8 processors and 11 communication links. Each node is a computing resource (a processor, or a cluster, or node).

A process can be in two states: running or blocked. In the running state (also called active state), a process has all the needed re and is either executing or is ready for execution. In the blocked state, a process is waiting to acquire some resource.

### B. The architecture

The target heterogeneous platform is represented by a directed graph, the platform graph.

There are p nodes $P_1, P_2,…, P_n$ that represent the processors. In the example of figure 1 there at eight processors, hence n = 8.

Each edge represents a physical interconnection. Each edge $e_{ij}: P_i \rightarrow P_j$ is labeled by value $c_{i,j}$ which represents the time to transfer a message of unit length between $P_i$ and $P_j$, in either direction: we assume that the link between $P_i$ and $P_j$ is bidirectional and symmetric. A variant would be to assume two unidirectional links, one in each direction, with possibly different label values. If there is no communication link between $P_i$ and $P_j$ we let $c_{i,j}= +\infty$, so that $c_{i,j} < +\infty$ means that $P_i$ and $P_j$ are neighbors in the communication graph.

### C. Wait – For – Graph (WFG)

In distributed systems, the sate of the system can be modeled by directed graph, called a wait for graph (WFG) [21,22,23,24,25]. In a WFG, nodes are processors and there is a directed edge from node $P_1$ to mode $P_2$ if $P_1$ is blocked and is waiting for $P_2$ to release some resource. A system is deadlocked if and only if there exists a directed cycle or knot in the WFG.

Let us first of all describe the deadlock condition problem more precisely.

A set $S = \{s_1, s_2,…s_k\} \subseteq \mathcal{E}$ of k > 1 entities is deadlocked when the following two conditions simultaneously hold:

Each entity $s_i \in S$ is waiting for an event permission that must be generated from another entity in the set;

No entity $s_i \in S$ can generate a permission while it is waiting.

If these two conditions hold, the entities in the set will be waiting forever, regardless of the nature of the permission and of why they are waiting for the "permission"; for example, it could be because $s_i$ needs a resource held by $s_j$ in order to complete its computation.

A useful way to understand the situations in which deadlock may occur is to describe the status of the entities during a computation, with respect to their waiting for some events, by means of a directed graph $\vec{W}$, called wait-for graph.

## IV. SOLUTIONS IN RESOURCE ALLOCATION HETEROGENEOUS DISTRIBUTED VIRTUAL MACHINES

In cloud computing model as introduced above, the resources provided is gathered in so many complicated steps. The development of a solution to prevent deadlock need to ensure that at least one of the following conditions cannot occur: Resources cannot be shared.Occupied and the additional resources required. No recovery resources. Exist in a cycle or knot.

### A. The proposed algorithm for distributing virtual machines

Virtual machine distribution on physical nodes at a specific time. To determine the distribution capabilities of all VM's of a lease on physical nodes at required times, starting at time t and lasting d seconds is  very difficult. When combining best-effort and  algorithm 2, we can find that the time before and algorithms used to provide resources in distributed environments is underutilized, as we cant schedule and best-effort request.

**Algorithm  1 Best-effort**
*Input: A lease l, a Boolean allow_future*
*Output:A lease l*
      $m \leftarrow map (l, now, duration[l])$
*Step 1 if $m \neq 0$ then*
        $VMrr \leftarrow new\ reservation$
        $start[VMrr] \leftarrow now$
        $end[VMrr] \leftarrow now + duration[l]$
        $res[VMrr] \leftarrow ..$
        add VMrr to reservations[l] and to slot table.
        $State[l] \leftarrow Scheduled$
*Step 2 else if m=0 and not allow_future*

*State[l] ←Queued*
*else changepoints ←t*
*For all cp ∈ changepoints do*
*m←map(l,cp,duration[l] )*
*if m≠ 0 then*
*break*
*end if*
*end for*
Step 3 return l

When the nodes have been sorted, the model uses the best-effort algorithm to distribute all VM's.

Aforementioned algorithm has the ability to distribute multiple VM's on the same node. With this research aiming to provide efficient distribution of resources, we propose the following technique in distributed environments. In this case, the algorithm tries to distribute as many VM's as possible on multiple physical nodes.

B. *The proposed for techique solution distributed environments*

When a lease (l) requests resources to create a virtual machine VM (including software, data operating systems, etc.) of any Data Center $DC_i$.

*Step 1 If $DC_i$ already has VM then $l_k$ already has resources, no deadlock detects and algorithm ends.*

*Else, if $DC_i$ does not have VM but l has been issued for transaction $l_j$ then send message $l_j$ block $l_k$ for $DC(l_j)$ and $DC(l_k)$. The message content is ($l_j$, $l_k$).*

*When any $DC_i$ received a notification message for blocked pair ($l_j$, $l_k$) then:*
*Step 2 If DC=DC($l_k$) then add $l_j$ to set P(DC) if $l_j$ does not belong to T(DC).*
*Step 3 If P(DC) ∩ T(DC) = {j} then deadlock detection succeeds and algorithm ends..*

*Else, send message ($l_j$, l') for all servers DC(l'), with each l' being a member of set B(S).*
*Else if DC≠DC($l_k$) then add $l_k$ to T(DC).*

From the above algorithm, it can be concluded that the proposed solution is of computational complexity. For every deadlock detection, the algorithm exchanges e request messages and e reply messages, where e=n(n-1) is the number of edges.

On consideration of resources required, we imposed some information about time that lease contracts are submitted, elapsed duration, the number of nodes required. We also set up more information: p = 1, m = 1024 (which means each node requires 1 CPU and 1024 MB of memory).

We conduct studies to evaluate the ability to provide resources, namely CPU hardware resources. In the future we will conduct additional analysis capabilities of virtualized resources such as storage drives, availability and completion time upon lease contract submission. p, the percentage of CPU being used by a given request. (Value of p can be 10%, 20%, 30%, 40% and 50% - because the percentage of CPU for using the greedy algorithm is calculated as approximately 49.20%).

VM, the number of nodes required. These are approximately as follows: small (1-24), medium (25-48), large (49-72). With the above two parameters, the research team determined the times to collect the results of the time when it requires to use the greedy algorithm within 1 lease contract in 1 DC and the time when the deadlock detection algorithm detected on deadlocks.

V. THE RESULTS FROM OUR ASSESSMENT

This section presents the results to the simulation experiments on simulated scheduling software Haizea.

**Table 1.** *Average time a contract ends with greedy algorithm, together with the CPU usage at local environment.*

| The ability of the CPU p | 10% | 20% | 30% | 40% | 50% |
|---|---|---|---|---|---|
| NOVM with NO suspension/recovery | 4,5 | 14,3 | 15,5 | 10,5 | 18 |
| NOVM with suspension/recovery | 2,4 | 2,6 | 6,5 | 16,5 | 20 |
| Can create VM | 2,8 | 2,8 | 27,3 | 37,3 | 90 |

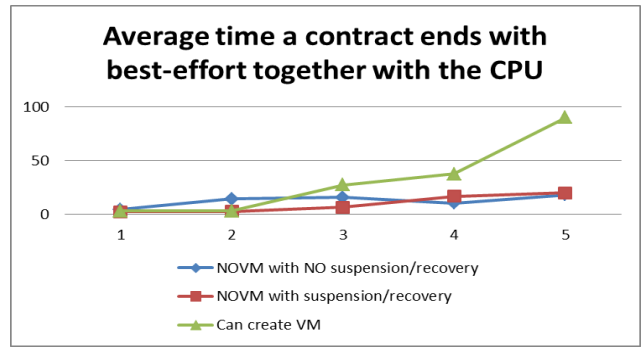

Figure 3. Graph showing the ability of CPU to each lease contract with best-effort algorithm

Through chart 1, by applying scheduling using greedy algorithm with the ability to provide resources for one lease contract (given the condition that the ability of CPU is pre-determined), we realized that the rates between failure and successful creation of virtual machines is the same. At CPU's ability of 50%, we can clearly see the that difference between these rates greater, with failure creation at 18 % and success creation at 90%.

**Table 2.** *Mean attenuation limit with more experiments*

| The ability of the CPU p | 10% | 20% | 30% | 40% | 50% |
|---|---|---|---|---|---|
| NOVM with NO suspension/recovery | 43,95 | 45,65 | 20 | 26,95 | 47 |
| NOVM with suspension/recovery | 4,05 | 8,40 | 8,46 | 4,05 | 8,40 |
| Can create VM | 45 | 43 | 66 | 75,5 | 85 |

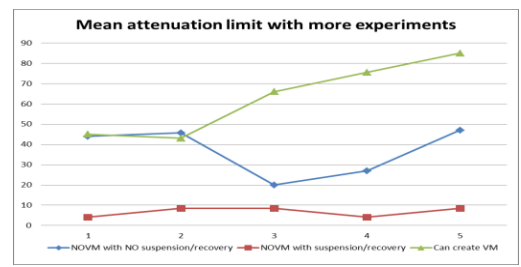

Figure 4. Graph showing mean attenuation limit after 10 trials, each lasted 180 minutes.

Through chart 2, by applying scheduling using greedy algorithm with the requirement to provide resources for 10 lease contracts (given the condition that the ability of CPU is pre-determined), we found that the success rate to create VM is high with the CPU's ability at 20%. As for CPU's ability at 10%, the success and failed rate are almost the same. At CPU's ability of 50%, we can clearly see that differences between these rates are greater, with failure creation at 47 % while success creation at 85%.

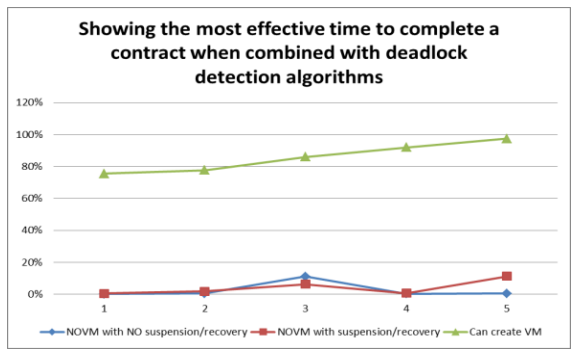

Figure 5. Graph showing lease contract completion time for each CPU capability in distributed environments, using deadlock detection algorithm.

Through chart 3, by applying scheduling using greedy algorithm combined with deadlock detection algorithm to the requirement to provide resources for 10 lease contracts (given the condition that the ability of CPU is pre-determined), we found that the success rate to create VM is very high with the CPU's ability at 50%. As for CPU's ability at 10% and 20%, the success creation is also higher than that of failure creation.

## VI. CONCLUSION

In the context of this paper, we are interested primarily in the criteria of readiness, because it affects preparing costs the most. The use of virtualization technology has great potential to meet the requirements of complex computing systems.

Two algorithms proposed in this research on providing efficient resources for virtual workspaces can grow up by utilizing the above advantages. Security problems, isolation, and the ability to adjust resources can impact positively on the standard of environmental quality by ensuring sufficient workspace resources (CPU, RAM, etc.) to support execution. Independence ability also improves the standard of resources openness, expanding pool of physical resources to run certain workspaces.

Our main approach focuses on applying scheduling algorithms for each type of lease contracts and applying the proposed algorithm in the distributed resources system. There were previous studies on the topic like that of author Borja Sotomayor, but it limited at researching local stations. We have also conducted experiments on distributed environments, given the ability of CPU, in some data centers – which yielded some positive results. It is the assessment that compares between the ability to create VM as requirements, or reject the request of creating a VM as other VM's cannot be suspended, or to stop the CPU in the data centers. Through this research we found that the application of appropriate scheduling algorithms would give optimal performance to distributed resources of virtual machine systems.

AUTHORS PROFILE

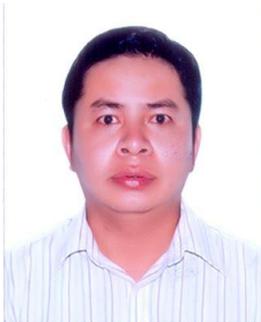

**HA HUY CUONG NGUYEN** received his B.S. degree in information technology from Van Lang University, Ho Chi Minh, Viet Nam, in 2003, the M.S. degree in Computer Science from DA Nan University, DA Nang, Viet Nam, in 2010. He was a lecturer, with Department of Information Technology, Quang Nam University, in 2003 so far. From 2011 until now, he studied at the center DATIC, University of Science and Technology - The University of Da Nang. At the center of this research, he doctoral thesis "Studies deadlock prevention solutions in resource allocation for distributed virtual systems". His research interests include network, operating system, distributed system and cloud computing.